\documentclass[fleqn,twoside,espcrc2]{article}
\usepackage{espcrc2}

\usepackage{graphicx}
\usepackage[figuresright]{rotating}

\newcommand{\AmS}{{\protect\the\textfont2
  A\kern-.1667em\lower.5ex\hbox{M}\kern-.125emS}}
\def\re#1{(\ref{#1})}
\def\beq{\begin{equation}}
\def\eeq{\end{equation}}
\def\beeq{\begin{eqnarray}}
\def\beeqn{\begin{eqnarray*}}
\def\eeeq{\end{eqnarray}}
\def\eeeqn{\end{eqnarray*}}

                 \def\D{\Delta}

\newcommand{\WW}{{\cal W}}

\newcommand{\no}{\nonumber}

\def\frac#1#2{ {{#1} \over {#2} }}

\def\ie{\hbox{\it i.e.}{ }}

\newcommand{\unity}{1\kern-.65mm \mbox{\form l}}
\newcommand{\ks}{\mbox{\form l}\kern-.6mm \mbox{\form K}}
\newcommand{\A}{A \raise0.5mm\hbox{\kern-1.8mm /}}
\def\pmb#1{\leavevmode\setbox0=\hbox{$#1$}\kern-.025em\copy0\kern-\wd0
\kern-.05em\copy0\kern-\wd0\kern-.025em\raise.0433em\box0}
\def\D{\hbox{\hbox{${D}$}}\kern-1.9mm{\hbox{${/}$}}}
\def\kbar{\hbox{$k$}\kern-0.2true cm\hbox{$/$}}
\def\nbar{\hbox{$n$}\kern-0.23true cm\hbox{$/$}}
\def\pbar{\hbox{$p$}\kern-0.18true cm\hbox{$/$}}
\def\nhbar{\hbox{$\hat n$}\kern-0.23true cm\hbox{$/$}}

\title{Area preserving diffeomorphisms and Yang-Mills theory in two
noncommutative dimensions \thanks{Talk delivered by A. Bassetto at ``Workshop on Light-Cone QCD and Nonperturbative
Hadron Physics 2005'', Cairns, Australia, July 2005.}}

\author{A. Bassetto\address[difipd]{Dipartimento di Fisica ``G.Galilei", 
	Via Marzolo 8, 35131 Padova, Italy\\ INFN, Sezione di Padova, Italy 
	\texttt{(bassetto, depol@pd.infn.it)}},
	G. De Pol\addressmark[difipd],
	A. Torrielli\address[humboldt]{Institut f\"ur Physik, Humboldt-Universit\"at 
	zu Berlin \\Newtonstr. 15, D-12489 Berlin, Germany 
	\texttt{(torriell@physik.hu-berlin.de)}}\thanks{Supported by DFG (Deutsche Forschungsgemeinschaft) within ``Schwerpunktprogramm Stringtheorie 1096''.},
	F. Vian\address[nordita]{NORDITA, Blegdamsvej 17, DK-2100 Copenhagen \O, Denmark 
	\texttt{(vian@nbi.dk)}}\thanks{Supported by INFN, Italy, and partially by the European 
	Community's Human Potential Programme under contract MRTN-CT-2004-005104~
	``Constituents, fundamental forces and symmetries of the universe''.}}
               
\begin{document}

\begin{abstract}
We present some evidence that noncommutative Yang-Mills theory in two dimensions
is not invariant under area preserving diffeomorphisms, at variance with the commutative case. Still, invariance under {\em linear} unimodular maps survives, as is proven by means of a fairly general argument.
\vspace{1pc}
\end{abstract}

\maketitle

\section{Introduction}
\noindent
Invariance under area preserving diffeomorphisms (APD) \cite{witten} is a basic symmetry of ordinary Yang-Mills
theories in two dimensions (YM$_2$). Thanks to this property, the theory acquires an almost topological
flavor \cite{m} and, as a consequence, can be solved. Beautiful group-theoretic methods
\cite{group} can be used to obtain exact expressions for the partition function and Wilson loop
averages. In particular, explicit solutions of the Migdal-Makeenko equation \cite{migmak} were obtained
in two dimensions \cite{kazakov} for  the expectation values of multiply-intersecting 
Wilson loops, due to the circumstance that they
depend only on the areas of the windows singled out on the manifold.

The invariance under APD was believed to persist in Yang-Mills theories defined on a noncommutative
two-dimensional manifold; actually it was believed to play a crucial role in the large gauge
group, characteristic of noncommutative theories, which merges internal and space-time
transformations. If this were the case one might expect to be able to solve noncommutative YM$_2$ by 
suitably generalizing the powerful geometric procedures employed in the ordinary commutative case.

This was suggested by an intriguing circumstance which occurs when studying 
the theory on a noncommutative torus.
Here one can exploit Morita equivalence in order to relate the model to its dual on a
commutative torus \cite{s} where the APD invariance is granted. The theory on the noncommutative plane
is then reached by a suitable limit procedure and one would expect invariance
to be preserved there \cite{ps2}.   

Wilson loop perturbative expansions in the coupling constant and in $1/{\theta}$, $\theta$ being
the non-commutativity parameter, were performed directly on the noncommutative plane
in \cite{bnt1,bnt2}. All the results obtained there, at the orders checked, were consistent with
APD invariance, the expressions depending solely on the area.

Only recently, the authors of \cite{adm} were able to extend to a higher order (the $\theta^{-2}$ term 
at ${\cal O}(g^4)$) those results, and found different answers for a Wilson loop on a circle and on a rectangle
of the same area. Their result motivated the systematic investigation presented in \cite{noi},
where Wilson loops based on a wide class of contours with the same area are considered.
The main issue of \cite{noi} is that, indeed, invariance under APD is lost even for smooth
contours, the breaking being rooted in the non-local nature of the Moyal product. Still
a residual symmetry survives, precisely the invariance under {\it linear} unimodular
transformations (SL(2,{\bf R})).

This report is devoted to illustrate the main points of \cite{noi}, which the reader is invited 
to consult for details and further references.

Perturbative evaluations of a Wilson loop, according to a well established procedure \cite{loop},
are most easily performed, in two euclidean dimensions, by choosing an axial gauge $n_{\mu}A_{\mu}=0$, 
$n_{\mu}$ being the (constant) gauge vector. As a matter of fact in such a gauge the self-interaction
terms are missing and no Faddeev-Popov determinant is required. Gauge invariance allows to transform
the gauge vector $n_{\mu}$ by a {\it linear} unimodular transformation $S_{\mu\nu},\, det S=1$,
with real entries, namely by an element of SL(2,{\bf R}). In turn this gauge transformation can be
traded for a corresponding {\it linear} area preserving deformation of the loop contour, as 
shown in \cite{noi}. In a noncommutative setting these transformations belong to the $U(\infty)$ gauge 
invariance group. The proof cannot be generalized to non-linear deformations, as they would
require a non-constant gauge vector, which would in turn conflict with the Moyal product. 

Since the considerations above seem to depend heavily on the axial gauge choice and gauge invariance
is explicitly {\it assumed}, it looks interesting to  prove the invariance under {\it linear} 
deformations of the contour {\it without changing the gauge vector}. This has been done for the $\theta^{-2}$ 
term in the expansion in $1\over \theta$ of the Wilson loop at ${\cal O}(g^4)$: it was indeed the term 
considered by the authors in \cite{adm}, where a difference was first noticed between Wilson loops of the same 
area but with different contours (a circular and a rectangular one). 
This quantity will in turn be used to prove the breaking of the {\it local}
unimodular invariance in the noncommutative context. 

The $\theta^{-2}$ term of the Wilson loop, at the perturbative order 
$g^4$, will be indicated as $W[C]$. Invariance would imply, for non 
self-intersecting contours,  that 
$W[C]=k A_C^4$, where $A_C$ is the area enclosed by the contour $C$,
the constant $k$  being universal, {\it i.e.} independent of the shape of $C$.

In the axial gauge $n A=0,\,$the non-planar contribution to the quantum average of the Wilson loop, at the lowest 
relevant perturbative order ${\cal O}(g^4)$, can be computed for a given contour $C$ parametrized by $z(s)\,,s\in [0,1]$
\mathindent=-20pt
\beeq
\label{order-4}
&&\WW_4^{np} = \int [ds] \, 
\tilde{n} \dot{z}_1
\ldots
\tilde{n} \dot{z}_4
\int \frac{d^2p\,d^2q}{(2\pi)^{4}}\times \\
&&\times \frac{\exp i \{ p\wedge q + p(z_1-z_3)+ q(z_2-z_4)\}}{(np)^2\,(nq)^2}
\,,\no 
\eeeq
\mathindent=0pt
 where $\tilde{n}$ is a unit vector orthogonal to $n$, and $p\wedge q$ is a shorthand for the antisymmetric bilinear
in the momenta $\theta p_{\mu}\epsilon_{\mu \nu} q_{\nu}$ ; two propagators correspond to four vertices on the 
Wilson line; and the peculiar phase factor $e^{i p\wedge q}\,,$ which originates from the noncommutative product, 
singles out the non-planar contribution. Performing the Fourier transforms in Eq.~\re{order-4} leads to complicate 
expressions: in order to study its physical content a further expansion, in powers of $1/\theta$, can be considered.
While the first terms turn out to depend 
only on the area $A_C$, the $1/\theta^2$ term is more involved and reads
\mathindent=-20pt
\beeq \label{makesing}
&&W[C]=\frac{g^4}{4! 4 \pi^2 \theta^2}  
{\mbox{\bf P}}\! \int\!\int\!\int\!\int dx_2 dy_2 dz_2 dt_2 \times\\
&& \times \frac{((x_1 -z_1)(y_2-t_2)-(y_1-t_1)(x_2-z_2))^4}{(x_2-y_2)^2(y_2-t_2)^2}
\,,\nonumber 
\eeeq
\mathindent=0pt
where the gauge $A_1=0$ has been chosen, the subscripts refer to the euclidean components of the coordinates, and the integral
is ordered according to $x<y<z<t\,.$ Eq.~\re{makesing} can be seen \cite{noi} to be equivalent to
\mathindent=-20pt
\begin{eqnarray} \label{makereg}
&&W[C]= \frac{g^4}{4! 4 \pi^2 \theta^2}\times \left[ A_C^4 +\right. \no \\
&&+ 30\, {\mbox{\bf P}}\!\int \!\int \!\int 
x_1 y_1 z_1 (x_1(y_2-z_2)+y_1(z_2-x_2)+ \no 
\\&&+z_1(x_2-y_2))\,dx_2 \,dy_2 \,dz_2 + \no \\ 
&&+\frac{5}{2} \oint \oint \left(\frac{4}{3}x_1^3 y_1 +x_1^2 y_1^2 +
\frac{4}{3}x_1 y_1^3\right)\times \no \\&& \times (x_2-y_2)^2\,dx_2 \,dy_2\,\left. 
\right]\,,
\end{eqnarray}
\mathindent=0pt
which is more convenient for our (both analytical and numerical) computations.
Here the triple integral is ordered according to $x<y<z$, while the double integral is not ordered.
If we define 
\beq\label{remnant}
W[C]=\frac{g^4 A_C^4}{4! 4 \pi^2 \theta^2} \, {\cal I}[C]\, ,
\eeq
it is apparent from dimensional analysis that ${\cal I }[C]$ is  dimensionless and  
characterizes the {\em shape} (and, {\it a priori}, the orientation) of a 
given contour. 

Invariance of the quantum average of a Wilson loop 
under translations is automatic in noncommutative theories owing to the trace integration over space-time
(see also \cite{abou}). In \cite{noi} the invariance of Eq.~\re{makereg} under linear area preserving 
maps (elements of SL(2,{\bf R})) is explicitly proven.

A basis for the infinitesimal generators of APD's is given by the set of vector fields
\mathindent=-20pt
\beeq
&&\{ V_{m,n}\equiv n x_1^m x_2^{n-1} \partial_{x_1}- m x_1^{m-1}x_2^n
\partial_{x_2}\,; \\&&(m,n)\in \mbox{\bf N}\times \mbox{\bf N} - (0,0)\}\,,\no
\eeeq
\mathindent=0pt
which form an infinite dimensional Lie algebra with commutation relations

\beq
\left[V_{m,n},V_{p,q}\right]=(np-mq)V_{m+p-1,n+q-1}\,.
\eeq
It follows that generators with $m+n\le 2$ span a finite subalgebra, corresponding to translations and linear 
unimodular maps. In \cite{noi} it is explicitely shown that Eq.~\re{makereg} is left invariant by this subalgebra.

In turn the breaking of the invariance under
{\it local}, non-linear area preserving maps is explicitly shown, at the perturbative level,
 for several different contours.

The computation of the Wilson loop is in principle straightforward for polygonal
contours, since only polynomial integrations are required; nonetheless,
a considerable  amount of algebra makes it rather
involved. 

Here we summarize our results:

\begin{itemize}

\item{{\bf Triangle:}} ${\cal I}[\mbox{Triangle}]$ was computed for an 
arbitrary triangle, the result being ${\cal I}[\mbox{Triangle}]=
\frac{8}{3}\simeq 2.6667$. This 
is  consistent with SL(2,{\bf R}) invariance, 
since any two given  triangles of equal area can be mapped into each
other  via a linear unimodular map.

\item{{\bf Parallelogram:}} ${\cal I}[\mbox{Parallelogram}]$ was computed for an 
arbitrary parallelogram
with a basis along $x_1$, the result being
${\cal I}[\mbox{Parallelogram}]=\frac{91}{36}\simeq 2.5278$. Again, this is 
consistent with SL(2,{\bf R}) invariance, by the same
token as above.

\item{{\bf Trapezoid:}} Here we can see analytically an instance of the
broken invariance.
Trapezoids of equal area cannot 
in general be mapped into one other by a linear 
transformation, since the ratio of the two basis $b_1/b_2$ is a SL(2,{\bf R}) 
invariant. 
One might say that the space of trapezoids of a given area, modulo  SL(2,{\bf R}),
has at least (and indeed, exactly) one modulus, which can  be conveniently 
chosen as the ratio $b_1/b_2$. Actually, the result we obtained for  
${\cal I}[\mbox{Trapezoid}]$ reads
\mathindent=-20pt
\beeq \label{trapezoid}
&&{\cal I}[\mbox{Trapezoid}]=
\\ &&=\frac{4(6 b_1^4 +24 b_1^3 b_2 +31 b_1^2 b_2^2 +
24 b_1 b_2^3 +6 b_2^4)}{9(b_1+b_2)^4}\,,\no
\eeeq
\mathindent=0pt
namely a function of $b_1/b_2$ only, duly invariant under the
exchange of $b_1$ and $b_2$, 
and correctly reproducing ${\cal I}[\mbox{Parallelogram}]$ when $b_1=b_2$, and 
${\cal I}[\mbox{Triangle}]$ when $b_1=0$. It is plotted in Fig.~\re{trap1} .

\begin{figure}[htb]
\includegraphics*[width=18pc]{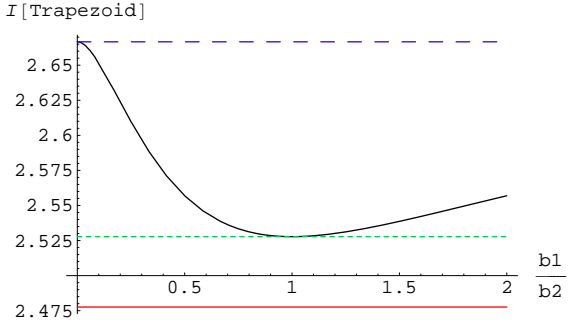}
\label{trap1}
\caption{${\cal I}[\mbox{Trapezoid}]$ as a function of the ratio of the basis
$b_1/b_2$; the continuous, dashed and dotted straight lines refer to
${\cal I}[\mbox{Circle}]$, ${\cal I}[\mbox{Triangle}]$ and ${\cal I}
[\mbox{Square}]$, respectively.}
\end{figure}

\end{itemize}

Thus, the main outcome of our computations is that different polygons
turn out to produce different results, unless they can be mapped into
each other through linear unimodular maps.

\smallskip

As opposed to polygonal contours, smooth contours  cannot be in general computed analytically, 
with the noteworthy exceptions of the circle and the ellipse. A circle can be mapped to any 
ellipse of equal area by the forementioned  area preserving linear maps. They share the result
\beq \label{circle}
{\cal I}
[\mbox{Circle}]={\cal I}[{\mbox{Ellipse}}]=1 +\frac{175}{12 \pi^2} \,.
\eeq

In the lack of explicit computations for smooth contours, different from 
circles and ellipses, this scenario might have left open the question whether 
in the noncommutative case invariance would still be there for smooth 
contours of equal area, only failing for polygons  due to the presence of 
cusps. We did the check by numeric computations  for the (even order) 
Fermat curves $C_{2n}\equiv \{(x,y):\ x^{2n}+ y^{2n}\ =1\,,$ {\rm with} 
$n\ge 1\}$,  which constitute a family of closed and smooth contours,
``interpolating'', in a discrete sense, between two analytically known 
results, \ie  the circle ($n=1$) and the square (the $n\to\infty$ limit).

\begin{table*}[htb]
\caption{Numerical results for ${\cal I}[C_{2n}]$ for Fermat curves
with different $n$.}
\label{tavola}
\newcommand{\m}{\hphantom{$-$}}
\begin{tabular}{@{}lllllllll}
\hline
$n$                 & \m1 & \m2 & \m3 & \m4 & \m5 &\m10 &\m15 &\m20\\
${\cal I}[C_{2n}]$  & \m2.4776 & \m2.4937 & \m2.5060 & \m2.5129 & \m2.5171 &\m2.5243 & \m2.5261 & \m2.5268 \\
\hline
\end{tabular}\\[2pt]
\end{table*}

\begin{figure}[h]{\includegraphics*[width=18pc]{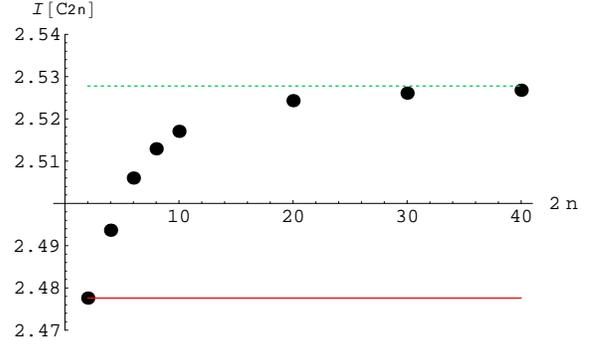}\label{fermatcurves}
\caption{${\cal I}[C_{2n}]$ for Fermat curves with different $n$; the
continuous, dotted lines refer to ${\cal I}[\mbox{Circle}]$, ${\cal
I}[\mbox{Square}]$, respectively.}}\end{figure}

${\cal I}[C_{2n}]$  definitely varies with $n$
and in the 
$n\to\infty$ limit approaches ${\cal I} [{\mbox{Square}}]$. 
Numerical results are given in Tab.~\re{tavola} and plotted in 
Fig.~2.
Thus we conclude that invariance under area preserving diffeomorphisms does 
{\em not} hold (see also \cite{cir-gri-sem-sza}).

\section{Beyond the perturbation theory}

Let us provide now a somewhat general argument concerning invariance under 
area preserving 
diffeomorphisms. We discuss for simplicity the $U(1)$ case.

In ordinary YM$_2$ the quantum average of a Wilson loop reads
\beq
\label{defi}
W[C]=\int {\cal D}\!A\,e^{-S[A]} w[C,A] \,,
\eeq
where $S[A]=\frac1{4}\int F_{ij}F_{kl}\eta^{ik}\eta^{jl} d^2 x$ is a functional of 
the vector field $A$, and $w[C,A]=P \exp{i\, \int_C A_i dx^i}$ is a functional of 
$A$ and of the contour $C$. 
This has been formulated in cartesian coordinates with metric $\eta_{ij}$.

Under a different choice of coordinates $x'=x'(x)$, $W[C]$ can be rewritten as  
\beq
\label{lop}
W[C]=\int {\cal D}\!A\, e^{-S_{gen}[A',\,g']} w[C',A'],
\eeq
where $S_{gen}[A,g]=\frac1{4}\int F_{\mu \nu} F_{\rho \sigma} g^{\mu \rho} 
g^{\nu \sigma} \sqrt{\det{g}} \,d^2 x$, 
provided $A$ and $\eta$ transform to $A',g'$ like tensors. Notice that $\det{g}$ is 
positive and the definition of $F_{\mu \nu}$ is left unchanged in the covariantized formulation.  

We can consider now the same functional computed for the deformed contour $C'$
\beq
\label{nlop}
W[C']=\int {\cal D}\!A\, e^{-S[A]} w[C',A],
\eeq
the deformation being described by the same map $x'=x'(x)$ as above.

The condition
\beq
\label{sym}
S_{gen}[A,g']=S[A]
\eeq
would describe a symmetry of the classical action.
In $d=2$, due to the circumstance that $F_{\mu \nu}$ is a two-form, we get 
\beq
\label{ac}
S[A]=\frac1{2} \int F_{12}^2 d^2x,
\eeq
while in $S_{gen}[A,g]$ the contractions with the inverse metric contribute a 
factor
$(\det g)^{-1}$
\beq
\label{gen}
S_{gen}[A,g]=\frac1{2} \int F_{12}^2 \frac{1}{\sqrt{\det{g}}}d^2x .
\eeq
The condition Eq.~\re{sym} then amounts to $\det{g}=1$, which is ensured if the 
Jacobian of the map is one,
namely if $C$ can be deformed to $C'$ by an area preserving map.

The classical symmetry persists at the quantum level if ${\cal D}\!A={\cal D}\!A'$,
apart from an overall normalization; this is suggested by the circumstance that the functional 
jacobian turns out to be independent of the fields.

\smallskip

In the noncommutative theory, the expectation value of the 
Wilson loop becomes 
\beq
\label{ncw}
W_{nc}[C,*]=\int {\cal D}\!A\, e^{-S_{nc}[A,*]} w_{nc}[C,A,*]\,,
\eeq
where the Moyal product
\beq
\label{Moy}
a * b \equiv \left[\exp{[i \frac{\theta}{2} \epsilon^{\mu \nu}
{\partial}^{x_1}_{\mu} {\partial}^{x_2}_{\nu}]} a(x_1) b(x_2)\right]_{|x_1=x_2}
\eeq
has been introduced, and the involved functionals are defined as
\beeq
S_{nc}[A,*]=\frac1{4}\int F_{ij}*F_{kl}\,
\eta^{ik}\eta^{jl} d^2 x \\
\nonumber
w_{nc}[C,A,*]=P_{*} \exp{i\, \int_C A_i dx^i}\nonumber \,,
\eeeq
in which the dependence on $*$ is explicitly exhibited.

$W_{nc}$ can be rewritten in general coordinates, provided a 
covariantized $*^{g}$ product is defined as
\mathindent=-20pt
\beeq
\label{gm}
&&a *^g b \equiv \\&&\left[\exp{[i \frac{\theta}{2} \frac{\epsilon^{\mu \nu}}{\sqrt{\det{g}}} 
{\cal D}^{x_1}_{\mu} {\cal D}^{x_2}_{\nu}]} a(x_1) b(x_2)\right]_{|x_1=x_2},\no
\eeeq
\mathindent=0pt
${\cal D}_{\mu}$ being the covariant derivative associated to the Riemannian 
connection for the metric 
$g$.~\footnote {We stress that, by introducing $*^g$, we are 
not formulating the theory on a curved space. Instead,
we are just 
rewriting the theory on the flat space
in general coordinates. It should be evident, from 
general covariance of the tensorial quantities involved, that
Eq.~\re{gm} 
in Cartesian 
coordinates reproduces the usual Moyal product Eq.~\re{Moy}. 
Notice also that,  
since by definition
${\cal D}_{\mu}g_{\rho \sigma}(x)=0$, it is irrelevant to choose either $x_1$
or $x_2$ as argument of $\det g(x)$ 
in Eq.~\re{gm}, and, by the same token, 
$*^g$ is uneffective 
when acting on the metric tensor  $g$.} 
 
Under the same reparametrization we then obtain
\mathindent=-20pt
\beeq
\label{rep}
&&W_{nc}[C,*]=\\
&&\int {\cal D}\!A\, e^{-S_{nc, gen}[A',g',*^{g'}]} w_{nc}[C',A',*^{g'}],\no
\eeeq
where the noncommutative action in general coordinates 
\beeq
\label{ncS}
&&S_{nc, gen}[A,g,*^g]=\\
&&\frac1{4}\int F_{\mu \nu} *^g F_{\rho \sigma} g^{\mu \rho} 
g^{\nu \sigma} \sqrt{\det{g}} \,d^2 x \no
\eeeq
\mathindent=0pt
and Wilson loop 
\beq
\label{Wil}
w_{nc,gen}[C,A,*^g]=P_{*^g} \exp{i\, \int A_{\mu}dx^{\mu}}
\eeq
 have been introduced. 

Assuming the absence of functional anomalies also in the noncommutative case,
we compare Eq.~\re{rep} 
with the 
functional $W_{nc}[C',*]$ computed for the deformed contour $C'$
\beq
\label{def}
W_{nc}[C',*]=\int {\cal D}\!A\, e^{-S_{nc}[A,*]} w_{nc}[C',A,*].
\eeq
The two quantities coincide if the following two sufficient conditions
are met
\begin{itemize}
\item $*^g=*$, 
\item
$S_{nc,gen}[A,g',*]=S_{nc}[A,*].$
\end{itemize}

These conditions imply that the map is at most linear, since the 
Riemannian
connection must vanish, and that 
its Jacobian 
equals unity. In conclusion, only SL(2,{\bf R}) linear maps are allowed.


\end{document}